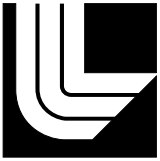

# FIRST-PRINCIPLES PHASE STABILITY, BONDING, AND ELECTRONIC STRUCTURE OF ACTINIDE METALS

P. Soderlind

March 14, 2013

**Disclaimer**

This document was prepared as an account of work sponsored by an agency of the United States government. Neither the United States government nor Lawrence Livermore National Security, LLC, nor any of their employees makes any warranty, expressed or implied, or assumes any legal liability or responsibility for the accuracy, completeness, or usefulness of any information, apparatus, product, or process disclosed, or represents that its use would not infringe privately owned rights. Reference herein to any specific commercial product, process, or service by trade name, trademark, manufacturer, or otherwise does not necessarily constitute or imply its endorsement, recommendation, or favoring by the United States government or Lawrence Livermore National Security, LLC. The views and opinions of authors expressed herein do not necessarily state or reflect those of the United States government or Lawrence Livermore National Security, LLC, and shall not be used for advertising or product endorsement purposes.

# First-principles phase stability, bonding, and electronic structure of actinide metals

Per Söderlind
Condensed Matter and Materials Division,
Lawrence Livermore National Laboratory

**Abstract**

The actinide elemental metals are scare, often toxic and radio active, causing challenges for both experiments and theory while offering fascinating physics. For practical purposes they are the prevalent building blocks for materials where nuclear properties are of interest. Here, however, we are focusing on fundamental properties of the actinides related to their electronic structure and characteristic bonding in the condensed state. The series of actinides is naturally divided into two segments. First, the set of lighter actinides thorium through plutonium, often referred to as the early actinides, display variations of their atomic volume reminiscent of transition metals suggesting a gradual occupation of bonding $5f$ states. Second, the heavier (or late) actinides, Am and onwards, demonstrate a volume behavior comparable to the rare-earth metals thus implying nonbonding $5f$ states. Arguably, one can distinguish plutonium metal as special case lying between these two subsets because it shares some features from both. Therefore, we discuss the early actinides, plutonium metal, and the late actinides separately applying first-principles density-functional-theory (DFT) calculations. The analysis includes successes and failures of the theory to describe primarily phase stability, bonding, and electronic structure.

## 1. Introduction

The series of actinide metals demonstrate behaviors unlike any other elemental solids in the Periodic Table. Properties such as equilibrium volume (or density), crystal structure, phase stability, thermal and electrical transport, anisotropy, among others, make these materials exotic in comparison to most metals. Because of their scarcity, nuclear instability, toxicity, and regulatory requirements, experimental work is generally challenging to say the least. Nevertheless, great strides have been made in recent years to illuminate their nature from various experimental angles [1].

Theoretically, the actinides pose challenge as well due to complexities of the electronic and crystal structures. Hence, the actinides have received increasing attention alongside technical developments of electronic band-structure methods [2] within the framework of DFT [3]. These techniques have since evolved to accurately account for relativistic effects including spin-orbit interaction [4] and complex crystal structures [5]. In recent years refinements of pseudopotential approaches, applying efficient plane-wave implementations, have demonstrated some success for actinides at least for the ground-state phases of Th-U [6].

The DFT band-structure approach is in principle only appropriate when dealing with bonding electrons that can be described by band states. In the early actinide metals this is indeed the case as we demonstrate in Figure 1. Here we plot the tabulated room temperature equilibrium volumes (full lines no symbols) for the $5d$ transition metals, $4f$ lanthanides, and $5f$ actinides [7] together with two opposing models of the $5f$ character. The $5f$ "fully bonding" model assumes band (delocalized) states for the $5f$ electrons while in the other, $5f$ "nonbonding", they are confined to core states with no band formation. Cleary, the former model is justified for the earlier actinides Th-Np with plutonium showing a slight deviation. At the same time, the nonbonding model reproduces the behavior of the heavier actinides Am-Bk. This result thus confirms the notion that the early actinides have bonding (itinerant) $5f$ electrons while the later do not. This perception [8] was founded on the fact that the $5d$ transition metals in Figure 1 behave as the early and the $4f$ lanthanides as the late actinides.

From Figure 1 it is evident that the early and late actinides fall into two categories with respect to the bonding character of their $5f$ electrons. One of them, plutonium, appears to not quite fit either simplistic model, particularly the δ-Pu (face-centered cubic, fcc) phase whose volume lies squarely in the middle between that predicted by the two bonding schemes. We have in the following chosen to divide the article into these segments, namely, the early actinides (Th-Np), plutonium, and the late actinides (Am-Cf), in Sections 3-5, after Section 2 that details with the computations. We further discuss the high-temperature body-centered cubic (bcc) phase that is prevalent in all actinides in Section 6 before some concluding remarks in Section 7.

## 2. Computational Details

All calculations are performed within the framework of DFT and the necessary assumption for the unknown electron exchange and correlation functional is that of the generalized gradient approximation (GGA) [9]. Although newer variations of this functional have been proposed the GGA remains the best choice for actinide metals [10]. Our particular implementation is based on the full-potential linear muffin-tin orbitals method (FPLMTO) that has recently been described in detail [11]. In addition to the choice of GGA, we have found that for actinides no geometrical approximations (full potential), full relativity including spin-orbit coupling, orbital polarization, and a well converged basis set is generally needed for good accuracy. Specifically, we associate a set of semi-core states $6s$ and $6p$ and valence states $7s$, $7p$, $6d$, and $5f$ to two kinetic energy parameters for a so-called double basis set. In all present calculations the sampling of k points in the Brillouin zone (BZ) for the appropriate summations are carefully checked for convergence. In the case of elastic-constant calculations, the converged number of k points in the irreducible BZ can sometimes be as high as several thousands but for most investigations a few hundred k points suffice. Sometimes specific calculations need further customization and we describe these in the applicable sections below.

## 3. Early actinides: thorium through neptunium

As one progress through the actinide series both the crystal structure as well as the electronic structure become more complex. The reason for this is that more $5f$ electrons are being involved in the bonding up until americium and they are occupying narrow bands close to the Fermi level [12]. At this point a change of the 5*f* character, akin to a Mott transition [8], takes place as we discussed in the introduction. Consequently, the early actinides are governed predominantly by fully active (itinerant) 5*f* bonding and the band-formation description is appropriate. Carefully performed electronic-structure calculations reproduce the correct ground state [12, 13] for these metals and also predict the correct structure-pressure relationships for Th, Pa, and U [14-17].

In Figure 2 we plot the body-centered tetragonal (bct) *c/a* axial ratio of thorium, theory and experiment, to highlight the accuracy of the DFT approach. The fcc to bct phase transition in Th was explained to be due to an increase of 5*f*-electron occupation with pressure [15]. Notably, in the case of Pa, calculations [16] predate experiment [18] by several years predicting a high-pressure orthorhombic phase (Cmcm, α-U-type.

A materials crystal structure is obviously sensitively connected to details of the bonding and electronic structure. Similarly, elastic constants characterize strength, anisotropy, and symmetry of the chemical bonds and constitute an accurate account of the bonding aspects of the electronic structure. In Table 1 we list calculated [19] elastic constants for α-U and contrast with experimental data. Notice that these compare as well as one can expect for simpler metals from the *d*-transition series (about a 20% difference between DFT and measurements). Other DFT calculations for α-U confirm this conclusion [20-22] suggesting that the results are robust and rather insensitive to technical approximations within the application of DFT.

From these examples we argue that the DFT band model is adequately describing the electronic structure of thorium, protactinium, and uranium. For neptunium metal the lack of experimental data makes this argument less convincing. However, the correctly predicted ground state (α-Np, primitive orthorhombic) and a good equation-of-state [23] suggest that the itinerant 5*f* electrons behave as described by the theory. DFT elastic constants for α-Np has recently been computed [24] and could in the future, when experiments become available, strengthen this proposition.

## 4. Special case: plutonium

Even though density-functional calculations were able to reproduce the ground state for plutonium, which is a non-intuitive complex monoclinic structure (α-Pu), a long time ago [25], the computations revealed issues that were not present for the earlier actinides. The obtained α-Pu equilibrium volume was somewhat small and the bulk modulus too large, but most serious, the description of the δ phase was completely wrong [25]. On the surface it thus appeared that DFT-GGA was only applicable for α-Pu and not any other of the many phases that exist, see the inset in Figure 3. Later [26] it became clear, however, that DFT-GGA was actually able to rather well reproduce most of the characteristics of the Pu phase diagram.

In Figure 3 we display calculated total energies for the known phases (compare the inset) and sensitive aspects such as hierarchy of the energies, their closeness, and atomic volumes, well reproduce the phase diagram. Even the calculated photoemission spectra is consistent with photoemission spectra, see the case for δ-Pu in Figure 4 [27]. The fact that the high-temperature ε (bcc) phase (Figure 3) appears somewhat high in energy is consistent with the findings for the earlier actinides and will be discussed separately below.

The undoubtedly intriguing consequence of the theory is the prediction of some form of magnetism, particularly for δ-Pu, a result that has never been experimentally verified [28]. The DFT-GGA model for δ-Pu suggests; (i) substantial (several Bohrs) spin and orbital magnetic moments, (ii) disordered magnetism (no long-range order), and (iii) an effective cancellation of magnetism due to anti-parallel spin and orbital moments of similar magnitude [27, 29]. Certainly, (ii) and (iii) make any experimental verification more challenging which may explain its elusiveness. Anti-ferromagnetic magnetism has, however, been detected in magnetic susceptibility measurements of $δ$-$Pu_{1-x}Ga_x$ alloy for as small x as 0.08 [30] and new experiments on plutonium have been proposed to seek the final answer on magnetism [31].

Consequently, the applicability of DFT-GGA for plutonium is controversial and many other models have been proposed to deal with the complexities of its electronic structure [32]. One argument for DFT is that structural and mechanical properties are well described within conventional DFT because the theoretical elastic constants for all its phases (except ε, see below) relatively well reproduce the experimental data [33]. In Table 2 single-crystal theoretical Pu elastic constants are averaged and related to polycrystal samples from resonant ultrasound spectroscopy (RUS) for all but the ε phase of Pu. Notice that the differences between the model and RUS are mostly less than about 20-30% and better when the RUS data are extrapolated to zero temperature for the high-temperature phases. This magnitude of discrepancy is similar to what has been found for transition metals [34] that do not involved uncertainties associated with the mentioned averaging schemes.

## 5. Heavier actinides: Am-Cf

As mentioned in the introduction, the ground-state phases of Am-Cf demonstrate localized 5$f$ electrons. As a result, the DFT-GGA band model cannot accurately reproduce the details of the electronic structure because of its bias towards delocalization for these materials in their ground state. This is evident from the calculated and measured electron spectra for americium, see Figure 5. Models that include stronger electron correlations, such as "DFT+U" and dynamical mean-field theory (DMFT) [35, 36] reproduce experimental spectra [37] far better [38]. Energetically, however, the DFT-GGA approach is still viable because the delocalized 5$f$ states can spin polarize which in terms of energetics and bonding captures much of the localization process [39].

One benefit the DFT-GGA approach has over the more strongly correlated methodologies is that it seamlessly deals with volume compression in the metal under pressure. High-pressure conditions (in the megabar range) will inevitably cause the 5$f$ states to form bands due to orbital overlap and in this regime DFT-GGA is entirely appropriate. Consequently, pressure-induced phase transitions can be most efficiently addressed [40] and indeed reproduce the results of modern diamond anvil-cell (DAC) experiments [41]. Figure 6 displays an example of this with calculated equations-of-state for Am parallel with DAC measurements and the agreement is clearly quite good.

Recently, curium metal has received attention experimentally and theoretically. It was found that some of the high-pressure Cm phases were stabilized due to magnetism [42]. In Figure 7 we display our own calculations that confirm the fact that CmIII (monoclinic C2/c) and CmIV (face-centered orthorhombic Fddd) phases can only be reproduced in spin-polarized calculations [43].

For some of the heavier actinides theory furthermore predicts new phases that are contrary to high-pressure measurements done earlier [44]. In Figure 8 and 9 we show the structural dependence on atomic volume for Bk, and Cf, respectively. Both metals are predicted to adopt a pressure-induced face-centered orthorhombic phase (Fddd) that is concurrent with a primitive orthorhombic phase (Pnma) in the case of Cf [45]. The previous experimental studies have instead promoted the orthorhombic α-U (Cmcm) phase [44] that is clearly too high in energy, see Figures 8 and 9, for both Bk and Cf. Because of the theoretical models ability to accurately describe the high-pressure behavior of americium [40], where more recent and more accurate experiments have been performed [41], we have greater faith in the theory for Bk and Cf and advocate new experiments.

## 6. Temperature stabilization of the bcc phase in actinides

So far, we have been discussing the successes and failures of the DFT-GGA approach for the actinide metals mostly considering their ground state. A much more difficult case to model, as it turns out, is the high-temperature bcc phase from which all actinides melts [46]. The ground-state theory predicts the bcc phase to be too high in energy, as mentioned above, and a more serious failure of the model is that bcc is *mechanically unstable* for, at least, the early actinide metals [47]. This is evident from the fact that the calculated tetragonal shear constant (C') is negative and in terms of modeling, this situation is a very poor starting point for development and improvements. One thus wonders if the bcc stabilization is due to strong electron correlation effects beyond that included in DFT-GGA, phonon interactions, or perhaps a combination of both. A clue is given by a DMFT investigation suggesting that strong electron correlations are not driving the bcc transition in plutonium [48].

Here we are examining the bcc phase for an actinide prototype, uranium (γ-U), with a relatively recent scheme to calculated finite temperature phonons from *ab initio* theory. The technique was originally devised to address the high-temperature bcc phase in the Group 4b metals Ti, Zr, and Hf. These all melt from the bcc phase while DFT at zero temperature predict the bcc phase to be unstable. For this purpose, the self-consistent *ab initio* lattice dynamics (SCAILD) method has been presented and utilized to show that for the Group 4b metals the bcc phase is stabilized due to phonon-phonon interactions [49]. The method can be adopted for the actinide metals but the scheme requires *ab initio* forces on the thermally displaced atoms and these are difficult to compute, particularly for actinides. We have previously reported on the details of these calculations for uranium [50] that are more accurate in their method of obtaining atomic forces than previously proposed [49].

In spite of the difficulties to calculate forces accurately [50], we are applying SCAILD for γ-U and in Figure 10 we show the phonon dispersions at 1113 K. Notice that all branches are positive thus inferring full mechanical stability that is due to the phonon-phonon interactions accounted for within SCAILD [49]. This result implies that the bcc phase is likely not a result of strong electron correlations simply because a weakly correlated model (DFT-SCAILD) is sufficient to achieve the stabilization. In Figure 11 we compare our calculated phonon density of states with neutron-scattering data at the same temperature [51]. The agreement is satisfying at higher energies but there is a systematic discrepancy at lower temperatures. We discussed [50] that the difference may be due to the nonlinearity of some phonon branches approaching the Γ point apparent in Figure 10.

Besides the phonons themselves, elastic constants can be extracted from the phonon dispersions and their slope approaching the Γ point. In Figure 12 we display phonon dispersions at (a) ambient-pressure volume, 20.86 Å$^3$, and (b) a compressed volume, 15.14 Å$^3$, where we also indicate the elastic-constant linear long wavelength limit. Due to the aforementioned deviation from linearity in some branches we are extrapolating from the linear part of these (dashed lines). The obtained elastic constants are collected in Table 3. Unfortunately, no experimental data are available to gauge the quality of our predicted γ-U elastic constants and we encourage experiments.

## 7. Summary

The phase stability, bonding, and electronic structure of a wide range of actinide metals are investigated by applying a parameter free DFT methodology. The actinides are here divided into the early and late actinides, and plutonium. The early actinides electronic structure is well described by DFT 5*f* band states and ground-state properties are robustly predicted. In the case of Pa theory predates experiments in the discovery of the high-pressure α-U phase, for example. The success of DFT for plutonium is more debated but we show that many facets of its phase diagram, which is highly nontrivial, is reproduced rather well. Magnetism is predicted while not evident in experiments so far and the model may be proven to be inaccurate in this regard.

As regards the late actinides, clearly the ground-state electronic structure is not correctly describing all aspects of the localized 5*f* electrons. We are making the argument, however, that properties derived from the chemical bonding are still relevant because the strong spin polarization effectively imitates localization. Under compression the 5*f* states become more band like and the DFT approach is increasingly accurate as exemplified in americium and curium. The model is also predicting new phases, distinct from early experimental suggestions, in berkelium and californium (Fddd and Pnma).

In conclusion, we stress that conventional DFT, if applied carefully, is very reasonable for actinides in general but with some obvious failures in regimes of strong electron correlation. Therefore, it is absolutely necessary to refine and develop methodologies that correctly deal with electron correlation in the strong as well as the weak limits.

**Acknowledgements**

We thank J. Tobin, A. Landa, L. Yang, O. Eriksson, and P. Souvatzis for enriching discussions. Computing support for this work came from the LLNL Computing Grand Challenge program. This work performed under the auspices of the U.S. DOE by LLNL under Contract DE-AC52-07NA27344 and funded by the Laboratory Directed Research and Development Program at LLNL under project tracking code 11-ER-033.

**Figure Captions**

1. Measured atomic volumes of the actinide metals (5*f*) are shown with a black line, a brown line for the 5*d* transition metals, and a green line for the lanthanides (4*f*). The red "5*f* nonbonding" and blue "5*f* fully bonding" curves show results from model calculations, assuming face-centered cubic structure, where the 5*f* electrons are treated as part of the valence band and localized to nonbonding core states, respectively.

2. Experimental data (open squares) and theoretical results (solid circle, solid line) for the *c/a* axial ratio in the tetragonal structure of Th as functions of pressure.

3. Total energies for plutonium metal. The inset shows the experimental volume-temperature phase diagram (Å$^3$ and K).

4. Comparison of experimental (dashed) and theoretical spectra, taken from [27].
5. Theoretical density of states for americium and experimental photoemission spectra, taken from [38].
6. Theoretical and measured [41] equation-of-state for americium, taken from [40].
7. Total energies, nonmagnetic and spin polarized, for CmIII (C2/c), CmIV (Fddd), and CmV (Pnma) phases of curium Taken from [43].
8. Total-energy differences for various phases relative to the bcc phase for berkelium. The orthorhombic (Fddd) phase is found in both Bk to be stable in a rather wide volume range.
9. Total-energy differences for various phases relative to the hcp phase for californium. The orthorhombic Fddd and Pnma phases are found to be concurrent around 50 GPa.
10. Calculated phonon dispersions for γ-U at 1113 K. Taken from [50].
11. Calculated and measured [51] phonon density of states for γ-U. Taken from [50].
12. Calculated phonon dispersions for γ-U at (a) 20.86 Å$^3$ and 1113 K, and (b) 15.14 Å$^3$ and 2000 K. The elastic-constant lines are marked at the Γ point. The transverse (T) Γ-H branches a nonlinear and an extrapolation is applied (dashed lines).

**Table Captions**
1. Calculated and measured single crystal elastic constants (GPa) for α-U. Taken from [19].
2. Bulk modulus (B), shear modulus (G), and $C_{11}$ = B+4/3G in units of GPa. Taken from [33].
3. Elastic constants for γ-U, obtained from DFT-SCAILD phonons (Figure 12).

**Tables**

**1.**

| Method | $c_{11}$ | $c_{22}$ | $c_{33}$ | $c_{44}$ | $c_{55}$ | $c_{66}$ | $c_{12}$ | $c_{13}$ | $c_{23}$ |
|---|---|---|---|---|---|---|---|---|---|
| DFT-GGA | 287 | 241 | 316 | 140 | 105 | 96 | 43 | 17 | 110 |
| Experiment | 215 (210) | 199 (215) | 267 (297) | 124 (145) | 73.4 (94.5) | 74.3 (87.1) | 46.5 | 21.8 | 108 |

**2.**

| Phase | Method | B | G | $C_{11}$ |
|---|---|---|---|---|
| α | DFT-GGA | 34.4 | 51.3 | 102.8 |
| α | Experiment | 37-54.4 | 43.5-43.7 | 104.6-112.8 |
| β | DFT-GGA | 38.5 | 25.3 | 72.2 |
| β | Experiment | 34.4 (41) | 18.2 (26) | 58.7 (75.7) |
| γ | DFT-GGA | 34.6 | 22.2 | 64.2 |
| γ | Experiment | 25.7 (31) | 16.5 (27) | 47.7 (67) |
| δ | DFT-GGA | 41.0 | 30.6 | 81.8 |
| δ | Experiment | 29.7 (38) | 16.2 (20) | 51.3 (64.7) |
| δ' | DFT-GGA | 44.0 | 31.4 | 85.9 |

**3.**

| Volume | Temperature | $c_{11}$ | $c_{12}$ | $c_{44}$ |
|---|---|---|---|---|
| 15.14 | 2000 | 570 | 390 | 90 |
| 20.86 | 1113 | 180 | 138 | 21 |

**Figure 1**

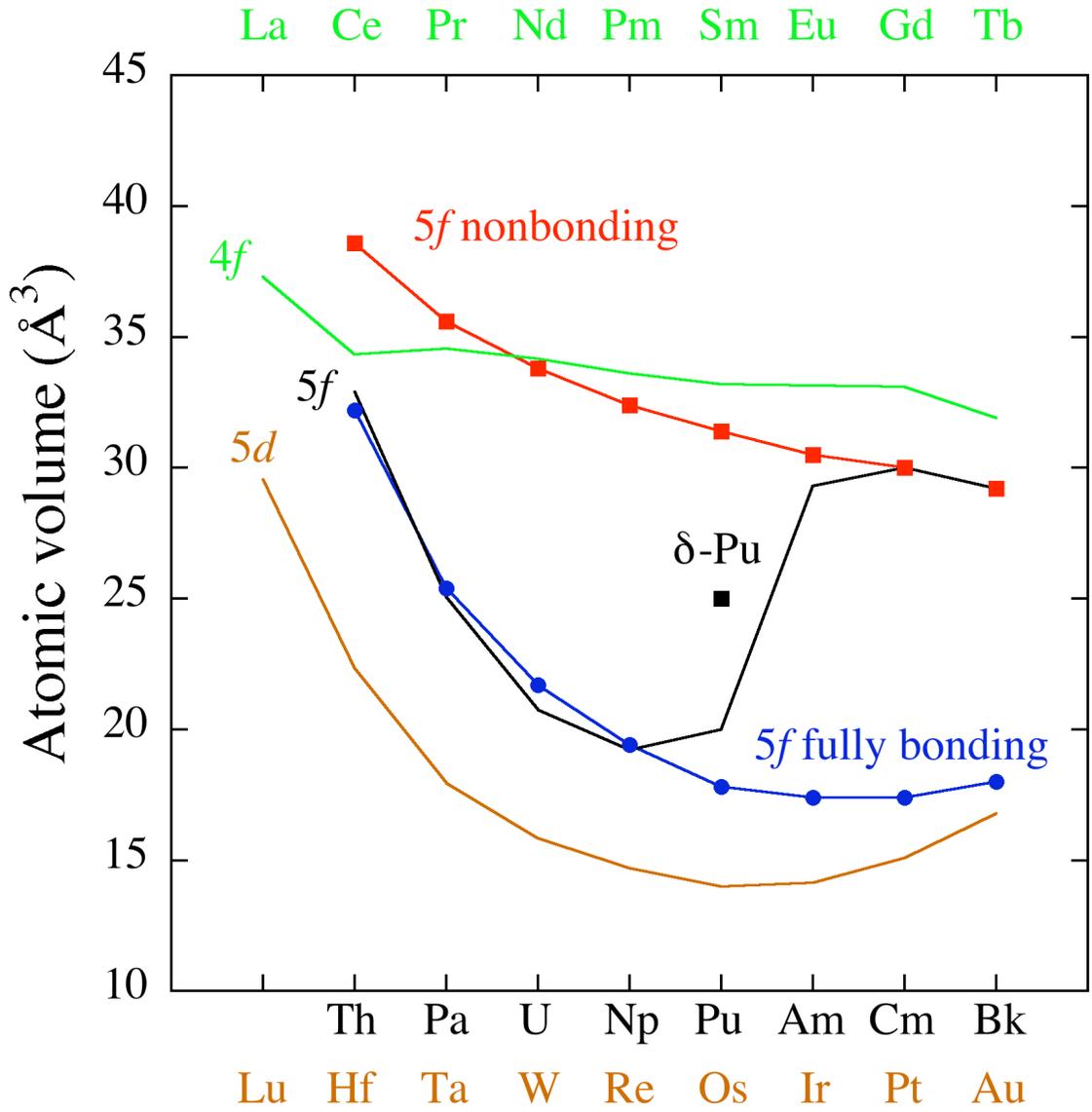

**Figure 2**

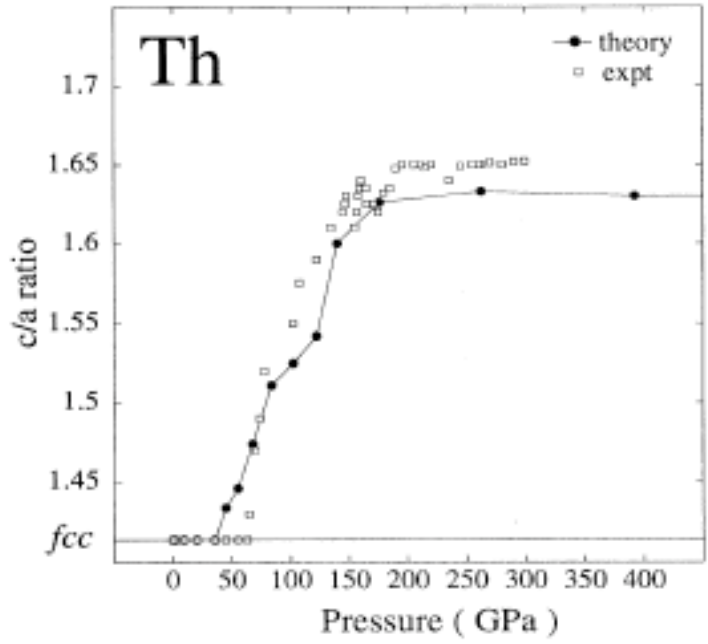

**Figure 3**

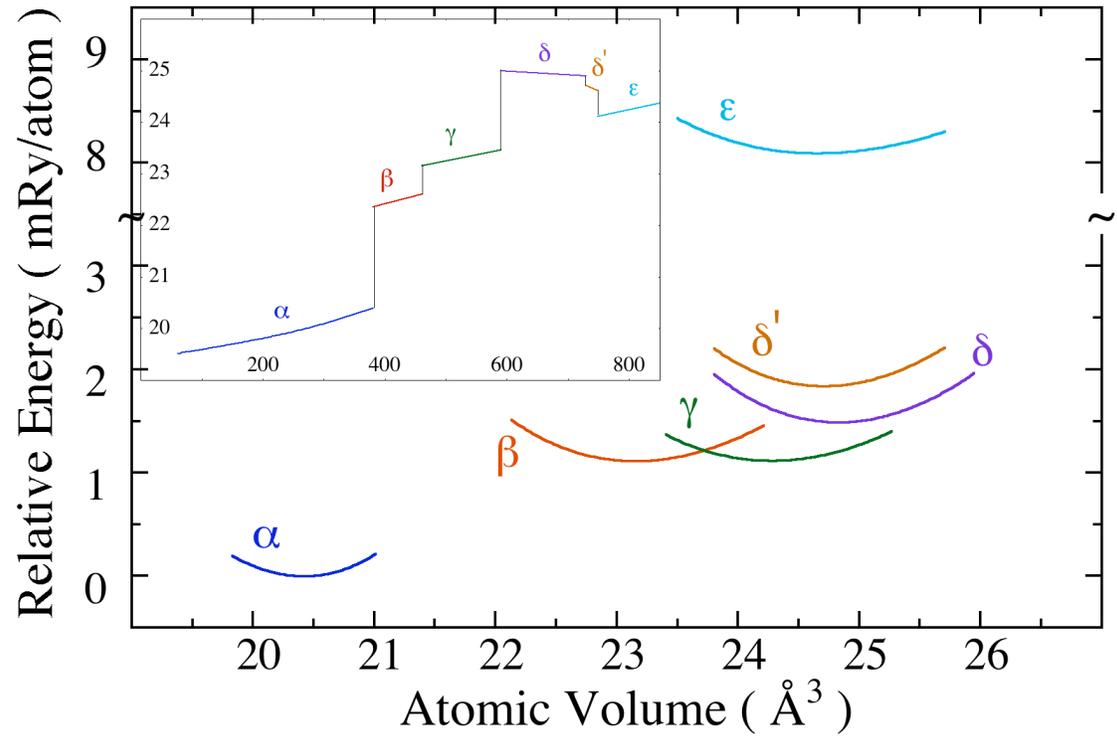

Figure 4

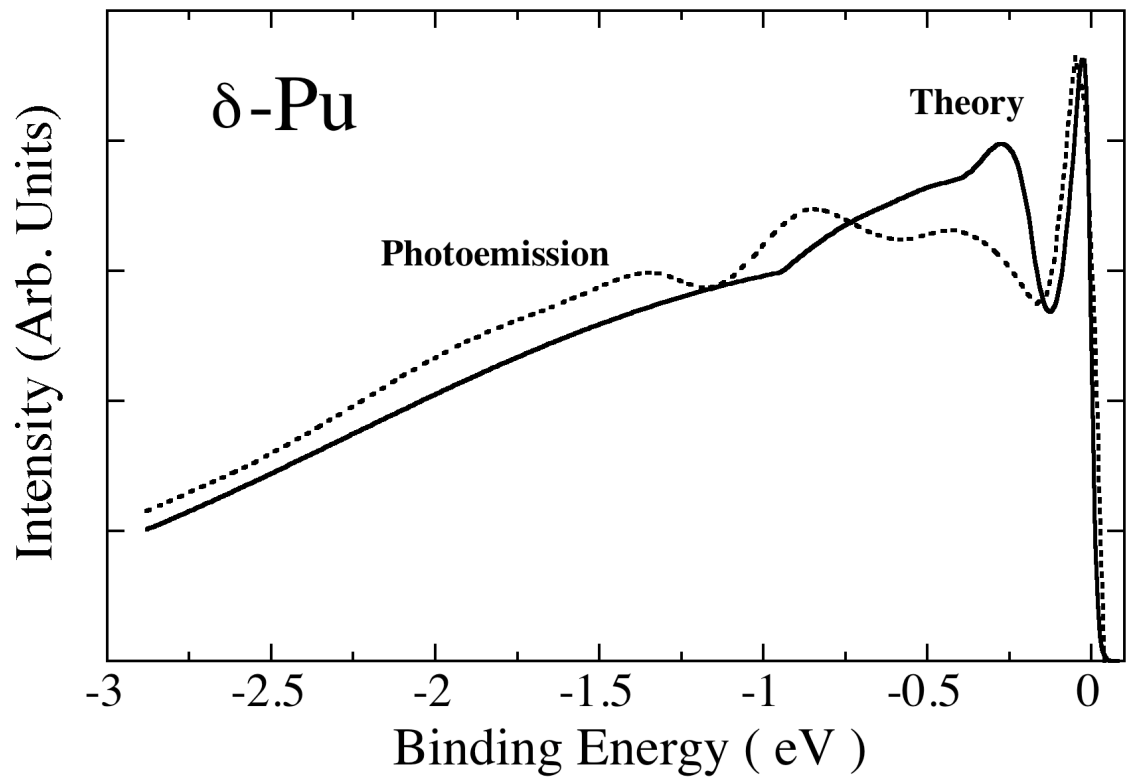

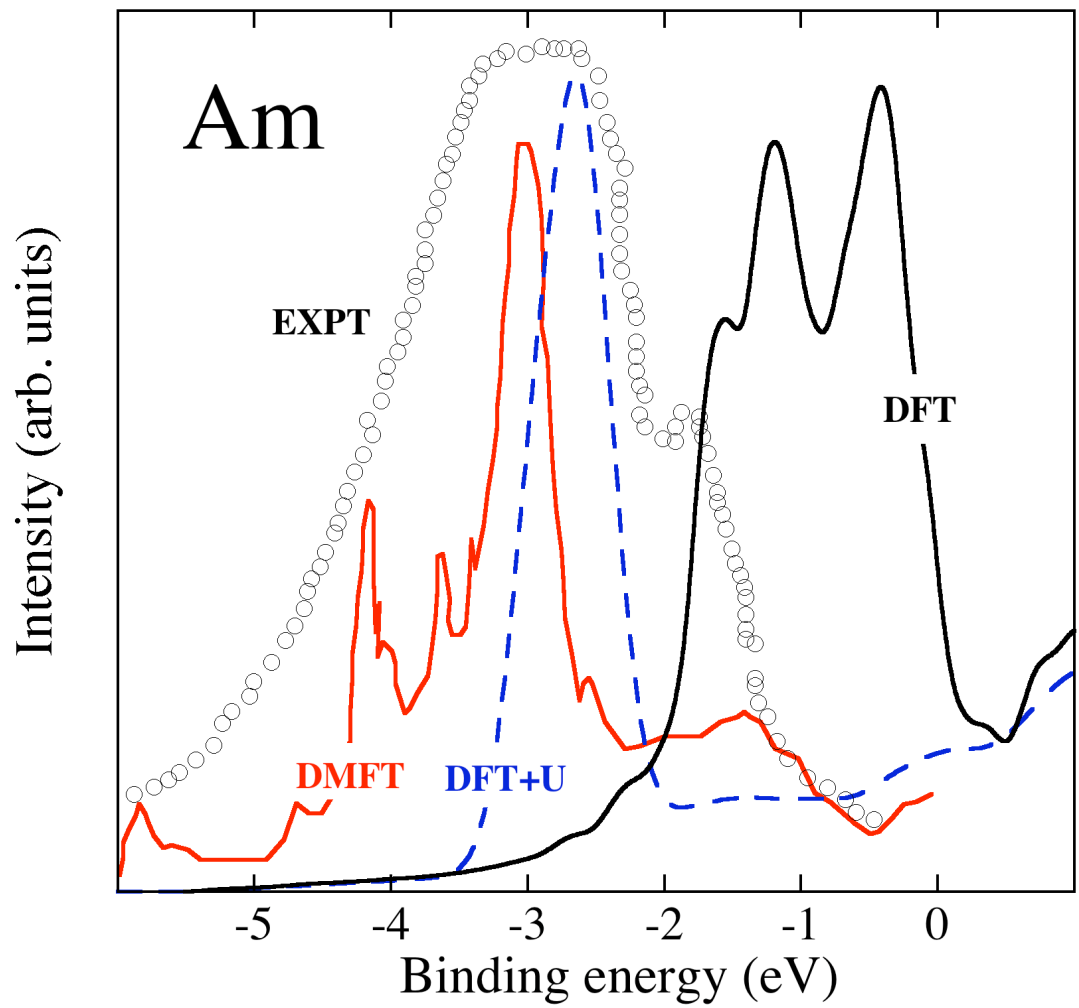

**Figure 6**

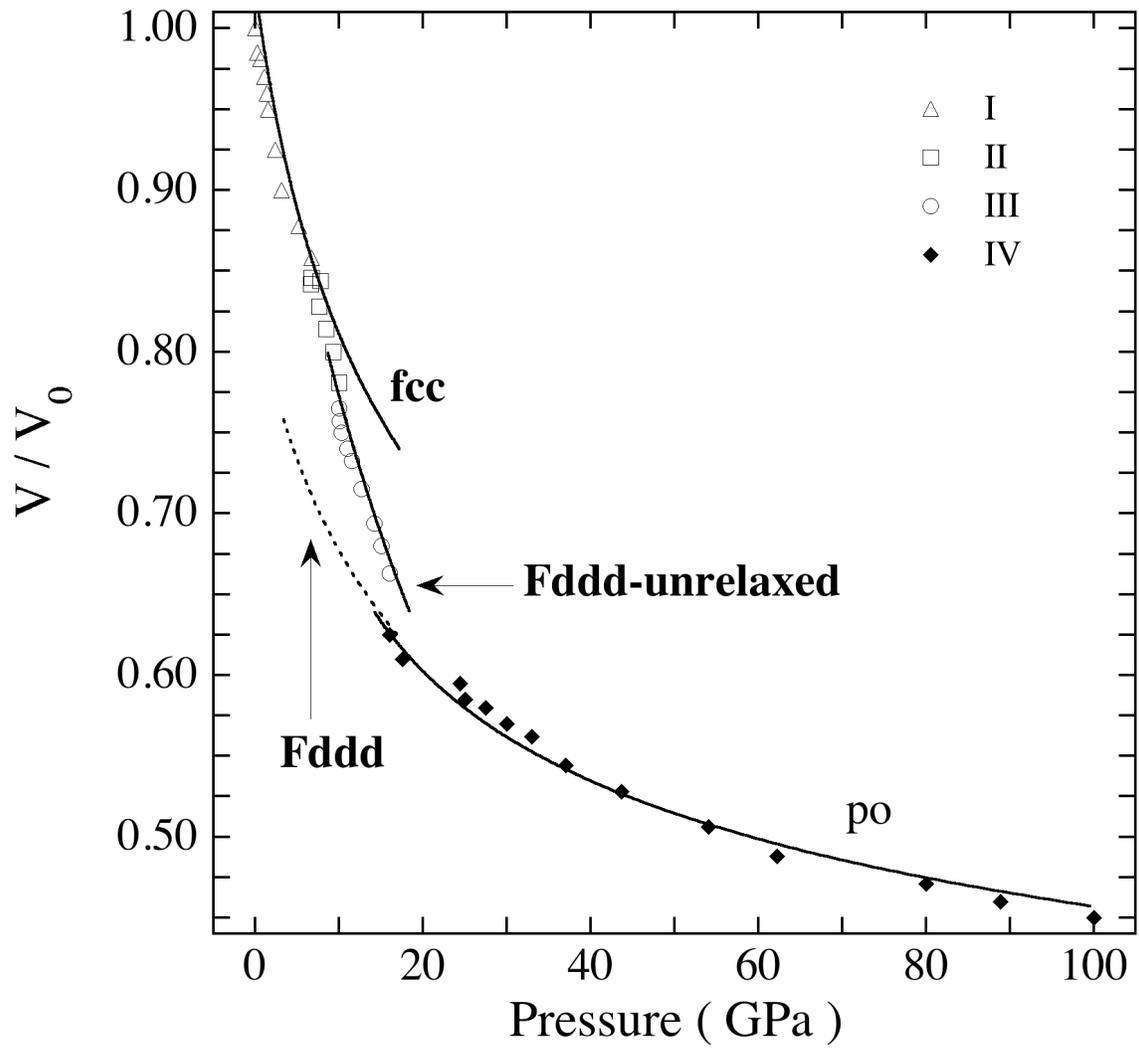

Figure 7

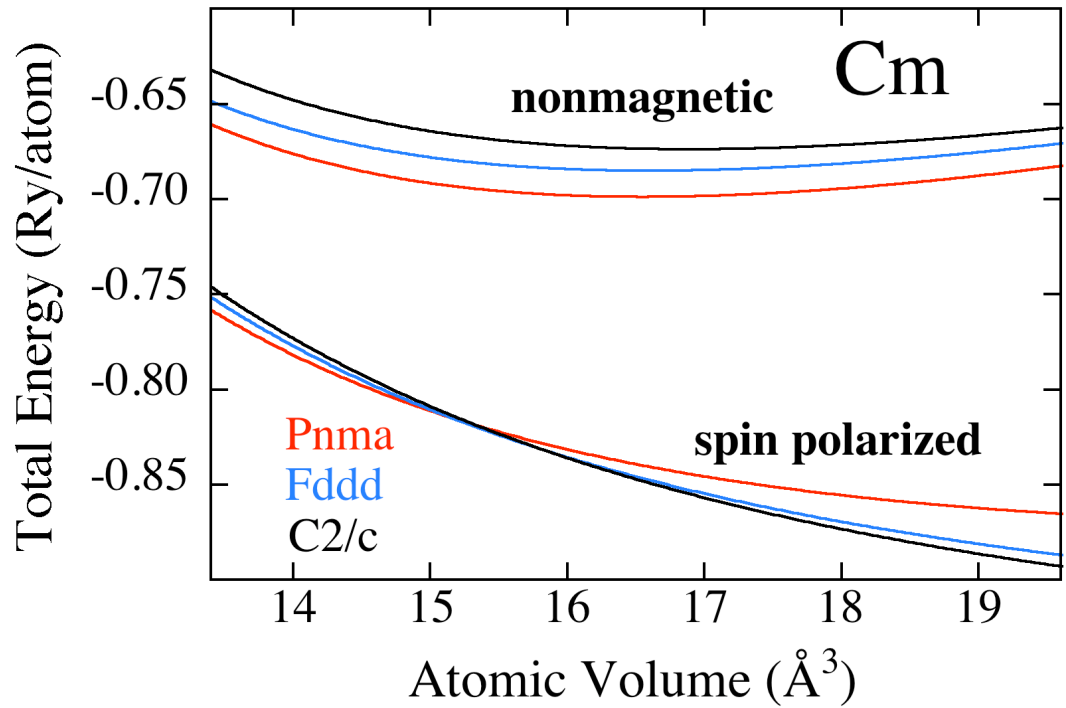

**Figure 8**

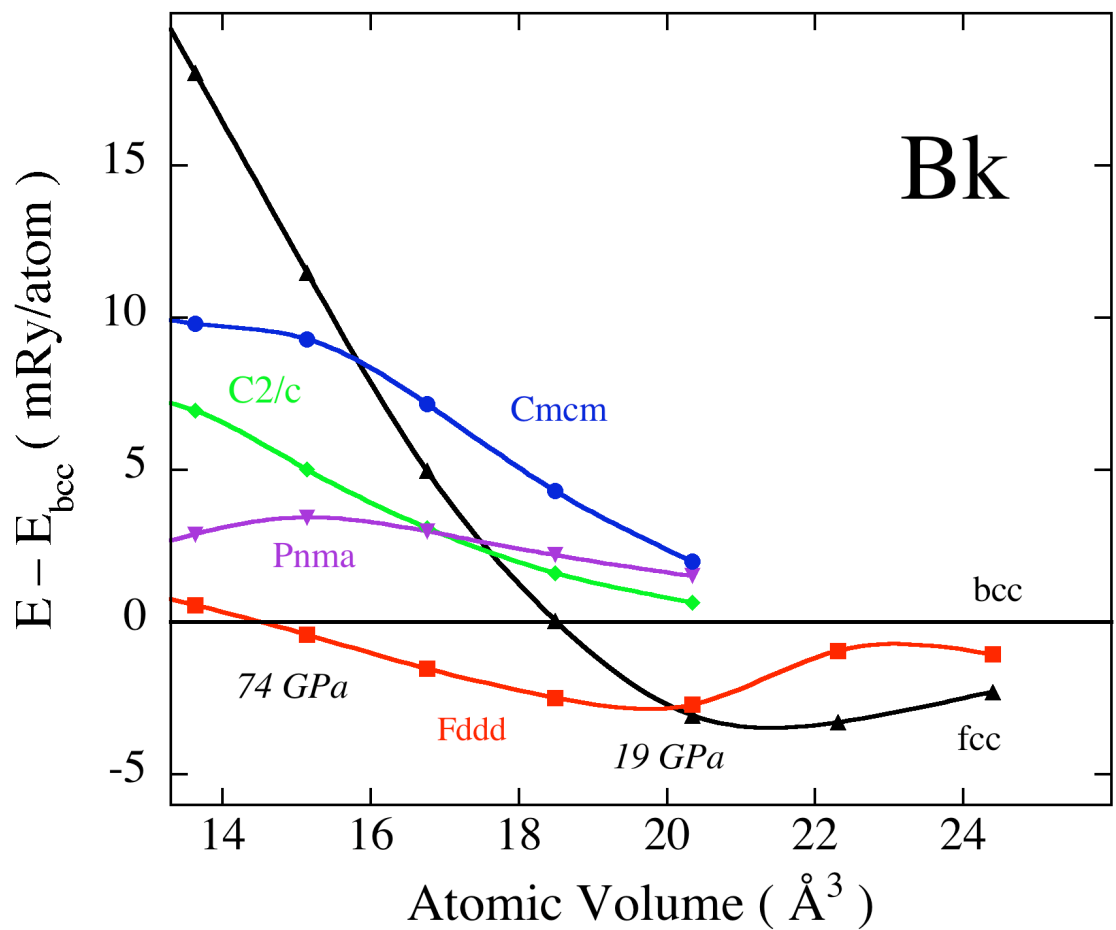

**Figure 9**

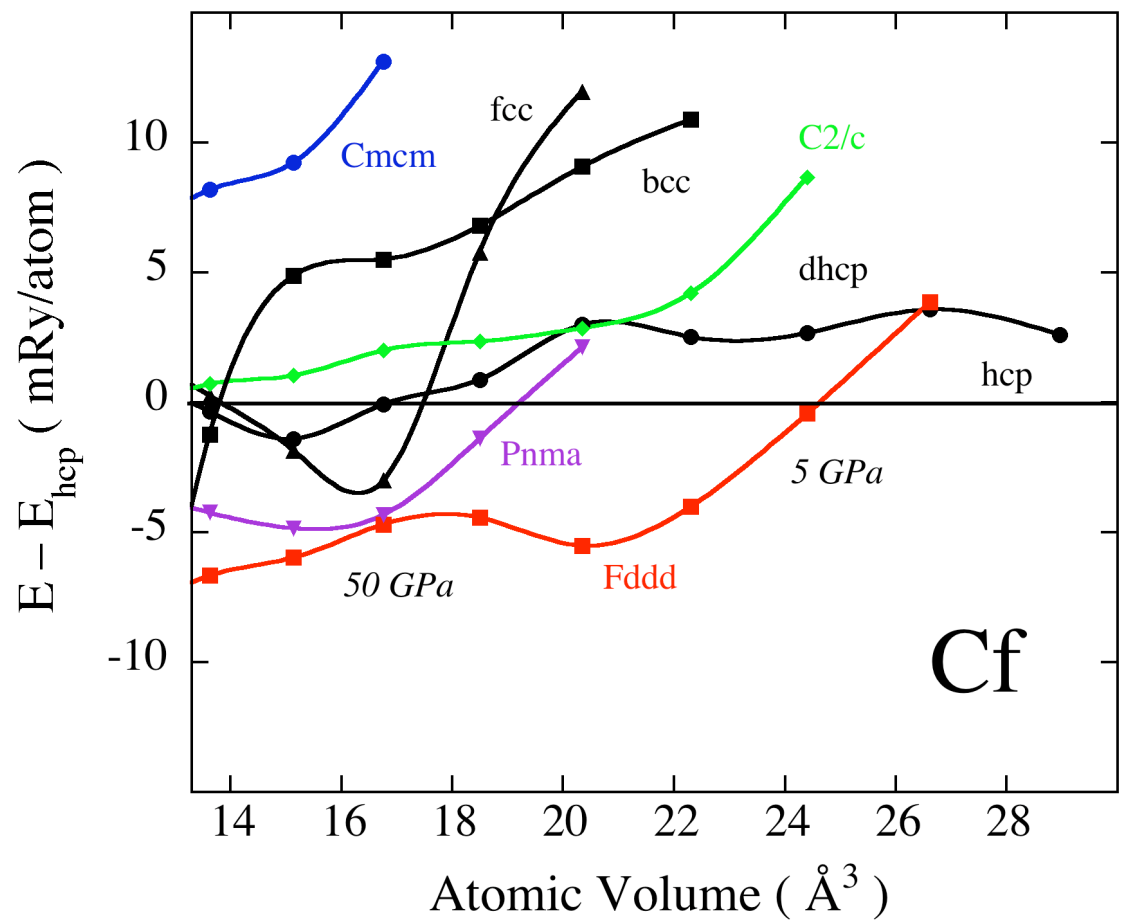

**Figure 10**

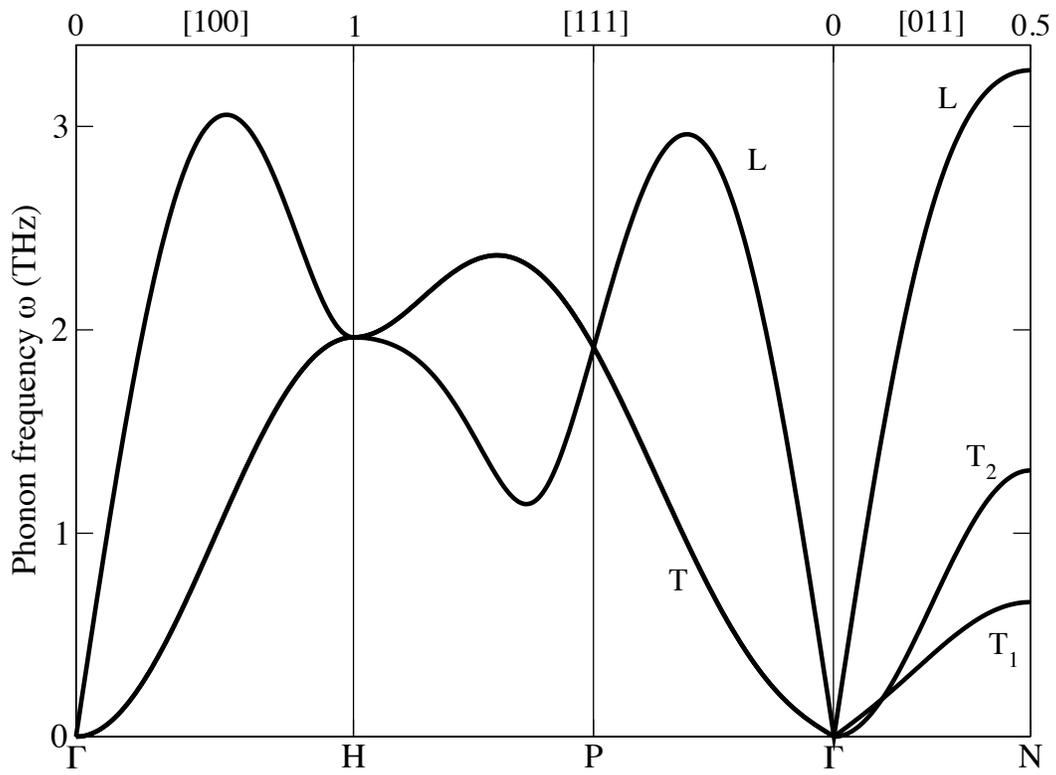

**Figure 11**

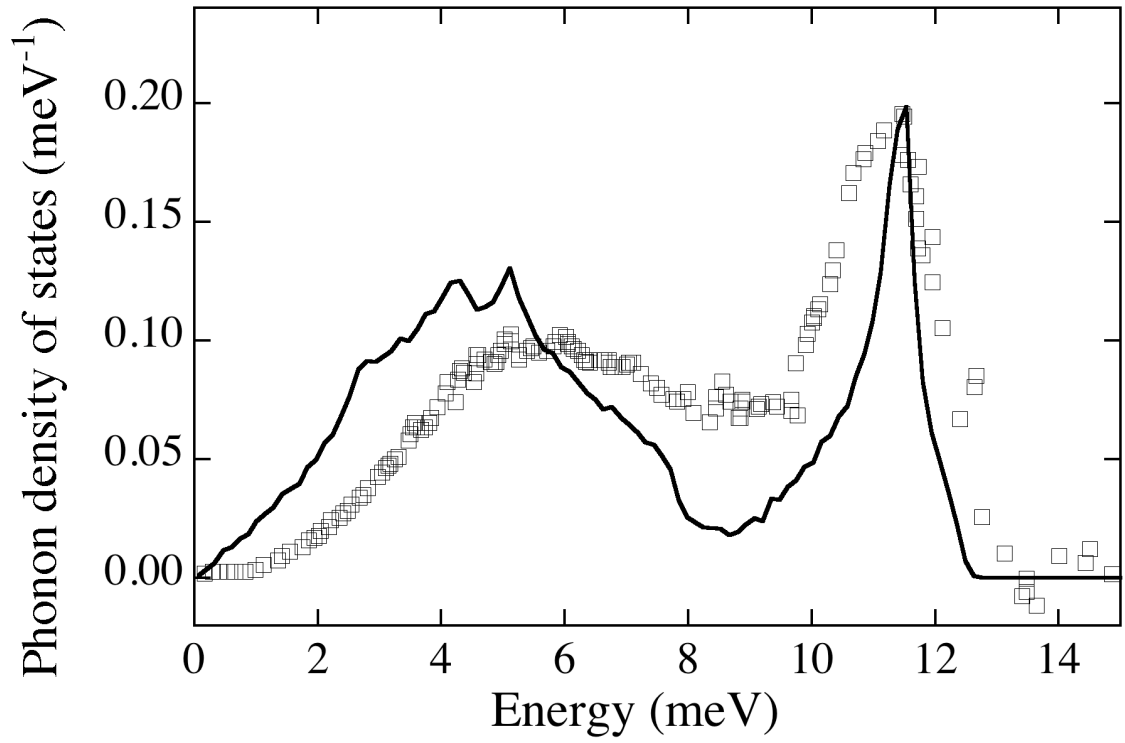

**Figure 12**

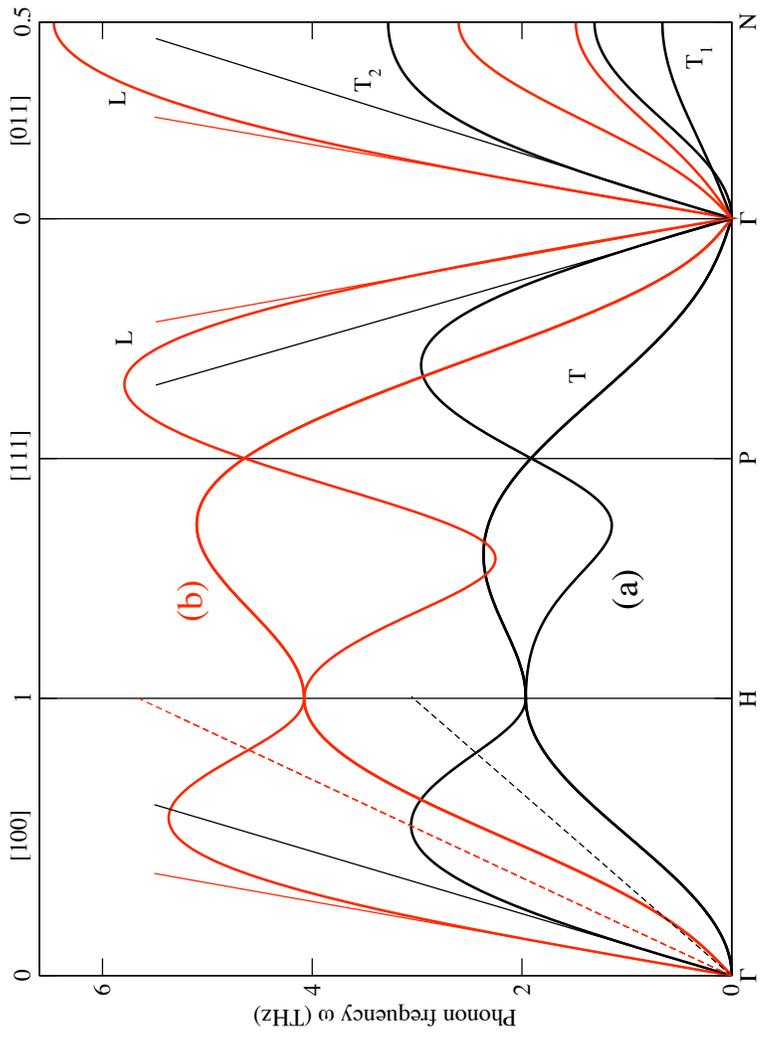